\documentclass{ws-p8-50x6-00}

\begin{document}

\title{Gauss-Bonnet interaction 
in Randall-Sundrum compactification 
\thanks{Talk at COSMO-2000, Cheju Island, Korea, Sep. 2000, based 
on works in collaboration with J. E. Kim and B. Kyae.}}
\author{Hyun Min Lee}
\address {Department of Physics and Center for Theoretical Physics,
Seoul National University, Seoul 151-742, Korea
\\E-mail: minlee@phya.snu.ac.kr}

\maketitle

\abstracts{
We show that the Gauss-Bonnet term is the only 
consistent curvature squared interaction in the Randall-Sundrum model and 
various static and inflationary solutions can be found.  
And from metric perturbations around the RS background with a single brane 
embedded, 
we also show that for a vanishing Gauss-Bonnet coefficient, 
the brane bending allows us to reproduce the 4D Einstein gravity at the 
linearized level.
}

\section{Introduction}
The main motivation of recent brane scenarios is to solve the gauge 
hierarchy problem in the higher-dimensional spacetime\cite{ADD,RSI}. 
In the supersymmetric  
Standard Model, the gauge couplings unify at the energy
 scale $\sim 10^{16}GeV $ by the renormalization group running from 
the weak scale. However, we cannot make such a prediction for the gravitational 
coupling, 
i.e., the Newton constant since gravity is not renormalizable. According to the Horava-Witten's proposal\cite{HW}, 
one finds that the 4D Planck scale 
becomes the low-energy arfifact of a four-dimensional world. In this proposal, 
the strong coupling limit of the $E_8\times E_8$ heterotic string compactified 
on a Calabi-Yau(CY) manifold $X$, is described by a 11-dimensional theory 
compactified on $X\times S^1/Z_2$ (the so called `Heterotic M-theory'). And the gauge and ordinary matter fields sit only on the 
ten-dimensional boundaries defined by $S^1/Z_2$, and gravity propagates in the 
bulk of the spacetime.
Upon the CY compactification of the Heterotic M-theory, the 
eleventh dimension is larger than the 
CY compactification length scale (or the string scale) when the 
string scale is identified as the GUT scale. Therefore, in fact, the universe 
is five-dimensional above the compactification scale of the eleventh dimension 
and thus smallness of the Newton constant stems from the fact that 
we cannot reach for the extra dimension of the universe. However, in the 
Horava-Witten's proposal, the five-dimensional(or 11-dimensional) fundamental
scale cannot be below the GUT scale for the validity of running of the gauge 
couplings. 
On the other hand, the higher-dimensional fundamental scale can be pulled down 
to the weak scale in the large extra dimension scenario suggested by 
Arkani-Hamed, Dimopoulos and Dvali(ADD)\cite{ADD}, where the higher dimensional 
spacetime is a factorizable geometry. 
In their proposal, at least two extra dimensions are 
required for solving the hierarchy problem. And, there is a problem related 
to the stabilization of the large extra dimensions, which corresponds to 
introduction of another hierarchy problem. 
By the way, the Randall-Sundrum(RS)'s proposal\cite{RSI} used the 5D 
non-factorizable  
geometry with one extra dimension to explain the gauge hierarchy problem. 
Their model setup  
is similar to that of the Horava-Witten's in the sense that the SM 
model matter and gauge fields are assumed to live only on the 4D boundaries (or 3-branes) 
defined by the $S^2/Z_2$ orbifold, but they introduced brane tensions on 
the boundaries and a non-zero bulk cosmological constant, which is shown to be 
realized from a Calabi-Yau compactification of 
the Heterotic M-theory\cite{CY}. As the physical scale varies along the bulk according to the exponential warp factor of the metric, they identify the 
positive (negative) tension brane as the hidden (visible) brane with the 
Planck (weak) scale. 
In this proposal, there is no large hierarchy between input parameters. 
In addition, the effective 4D Planck scale becomes still finite even for 
the infinite extra 
dimension, which implies an alternative compactification without small extra 
dimension\cite{RSII}. 
If all mass scales in the RS model are given by one input scale, i.e., the 5D 
Planck scale, then the curvature scale is also of order of the 5D Planck scale.
 Therefore, the next step is to consider the higher order gravity effects in the
RS model. 
\\ 
\indent
In this paper, we show that in the existence of the Gauss-Bonnet term, 
various static and inflationary solutions can be found and properties and 
pecularities of the RS model can be maintained\cite{KKL}. Then, 
through the perturbative analysis and the brane bending effect, we consider 
the second RS model with the Gauss-Bonnet term as a 4D effective gravity 
theory\cite{KL}.

\section{A review of RS model}
The large extra dimension scenario\cite{ADD} is the simpliest case to 
use the higher-dimensional mechanism to solve the gauge hierarchy problem. 
The effective 4D Planck scale $M_P$ is determined by the (4+n)-dimensional
Planck scale $M$ and the geometry of the extra dimensions.
Since the higher-dimensional spacetime is a product of a 4-dimensional 
spacetime with a n-dimensional compact space in the large extra dimension 
scenario, the effective 4D Planck scale $M_P$ is given by the formula 
$M^2_P=M^{n+2}V_n$, where $V_n$ is the volume of the compact space. For the 
(4+n)-dimensional Planck scale $M$ to be the weak scale, the compactification 
scale $\mu_c\sim 1/V^{1/n}_n$ would have to be much smaller than the weak 
scale, which requires that the SM particles and forces are confined to a 
4-dimensional 
subspace while gravity is allowed to propagate in the bulk of the spacetime. 
However, it gives rise to a new hierarchy problem related to the 
compactification scale.
On the other hand, the small extra dimension scenario\cite{RSI} considers 
the higher-dimensional spacetime as the case of the   
non-factorizable geometry with $S^1/Z_2$;
\begin{eqnarray}
ds^2&=&e^{-2kb_0|y|}\eta_{\mu\nu}dx^\mu dx^\nu+ b^2_0 dy^2 \nonumber \\
&\equiv&\bar{g}_{MN}dx^M dx^N\label{rsm}
\end{eqnarray}  
where $k$ is the AdS curvature scale given by $k=\sqrt{-\frac{\Lambda_b}{6M^3}}$ and $y$ is the fifth coordinate with $y\in [-\frac{1}{2},\frac{1}{2}]$. 
Then, the effective 4D Planck scale is determined by $M^2_P=M^3\int_{-1/2}^{1/2}dy \, e^{-2kb_0|y|}=\frac{M^3}{k}(1-e^{-kb_0})$, which implies the weak 
dependence of the 4D Planck scale on the extra dimension. Even the non-compact 
extra dimension also allows the finite 4D Planck scale. Since there exists 
two 3-branes with brane tensions $\Lambda_1$, $\Lambda_2$ at $y=0$ and 
$y=\frac{1}{2}$, respectively, by the consistency of the boundary 
conditions on the branes, the following finetuning 
condition is required between brane and bulk cosmological constants,
\begin{eqnarray}
\Lambda_1=-\Lambda_2=\sqrt{-6M^3\Lambda_b} .\label{finetune}
\end{eqnarray} 
And, as the warp factor exponentially decreases along the bulk, we can obtain 
the weak scale as the physical scale of the brane at $y=\frac{1}{2}$ by      
appropriately choosing the distance between two branes 
(i.e., $b_0\sim 74/k\sim 74/M_P$) without introducing another large hierarchy. 
\\
\indent
In a next step, cosmological considerations in the extra dimension scenarios 
follow essentially. The cosmological bound on the ADD scenario                 
comes from the effects of the light Kaluza-Klein(KK) graviton excitations. 
However, the masses of the KK graviton modes should be larger than about a few 
GeV for nuclesynthesis\cite{Chang}, which corresponds to $b_0< 80/k$, so it 
seems that the light KK graviton problem may be avoided in the RS model. On the 
other hand, we have to consider the cosmological expansion of 3-brane 
universes in the bulk and check whether the normal Hubble expansion rate can be 
reproduced on the brane. For the sake of this, we assume that the 3-branes are
homogeneous and isotropic such that the 5D metric reads,
\begin{eqnarray}
ds^2=-n^2(\tau,y)d\tau^2+a^2(\tau,y)\delta_{ij}dx^i dx^j +b^2(\tau,y)dy^2
\label{metric}.
\end{eqnarray} 
Then, from the Einstein equations of motion with the above metric, we have the  
following non-trivial equation for the Hubble expansion rate on 
each brane\cite{cosm},
\begin{eqnarray}
H^2&=&\frac{\Lambda_b}{3M^3}+\frac{(\rho_i+\Lambda_i)^2}{36M^6}+\frac{K}{a^4}|_i\nonumber \\
&=&sgn(\Lambda_i)\bigg(\frac{\rho_i}{3M^2_P}\bigg)+\frac{\rho^2_i}{3|\Lambda_i|M^2_P}+\frac{K}{a^4}|_i, \,\,\,\, i=1,2
\end{eqnarray}
where we used the Eq.~(\ref{finetune}) and $K$ is a constant of motion 
determined from the initial condition and the last term is so called the dark 
radiation\cite{drad}. Consequently, the $\rho^2_2$ term and the dark radiation 
term in the above Hubble expansion would drastically affect a later cosmology 
in our brane, e.g., the big bang nucleosynthesis and there is the wrong sign 
problem in the Hubble parameter from the linear term in $\rho_2$.
To avoid the 
effects to the nucleosynthesis, the brane tension must be $|\Lambda_2|\gg(MeV)^4$ and the dark radiation density should be diluted by
inflation and/or reheating processes. And the wrong sign problem can be solved 
by having the positive tension brane $\Lambda_2 >0$\cite{KKL} or by introducing a mechanism for stabilizing the size of the extra dimension while the 
branes expand\cite{GW,radius}.

\section{Static and inflationary solutions}
When the higher curvature terms are added as correction terms in the action, 
the higher derivatives are generically induced in the equations of motion, 
which gives rise to runaway solutions and tends to make the system unstable. 
In particular, since the first derivative of the RS metric has to be 
discontiuous along the bulk to 
compensate the delta function sources due to the branes, 
we have to choose the higher curvature terms such that there don't appear 
higher derivatives of the metric with respect to the $y$ coordinate than 
the second. The Gauss-Bonnet term, $E=R^2-4 R_{MN}R^{MN}+R_{MNPQ}R^{MNPQ}$, 
one of particular choices of the curvature
squared terms, is a topological term in $D=4$ and it does not affect the 
graviton propagator even for the $D>4$ flat spacetime background\cite{ZZ}. 
Since there
are no higher order derivatives induced from the Gauss-Bonnet term, it seems
that the Gauss-Bonnet term is consistent with the RS model as the effective 
interaction. 
\\ 
\indent
When the Gauss-Bonnet term is included as the effective interaction in 
the RS model, we obtain two RS type static
solutions with the AdS curvature scale $k$ as follows\cite{KKL},
\begin{eqnarray}
k=k_{\pm}\equiv \bigg(\frac{M^2}{4\alpha}\bigg[1\pm 
\sqrt{1+\frac{4\alpha\Lambda_b}{3M^5}}\bigg]\bigg)^{1/2} \label{curvature}
\end{eqnarray}
where $M$, $\alpha$ and $\Lambda_b$ are the 5D fundamental scale, 
the dimensionless parameter of the Gauss-Bonnet term and the bulk 
cosmological constant, respectively.  By the boundary conditions on 
the branes, the finetuning conditions are to be  
satisfied between input parameters, 
$\alpha$, $\Lambda_1$, $\Lambda_2$ and $\Lambda_b$\cite{KKL},
\begin{eqnarray}
\Lambda^\mp_1=-\Lambda^\pm_2 
=\mp6k_\pm M^3\sqrt{1+\frac{4\alpha \Lambda_b}{3M^5}}\label{gbfine}.
\end{eqnarray}  
From the above result, we find that for the $k_+$ solution,   
the bulk cosmological constant is allowed to be positive and it is possible 
to have a positive tension brane as the visible brane at $y=\frac{1}{2}$. 
On the other hand, the $k_-$ solution is connected with the RS solution in the
limit of vanishing $\alpha$, for which the visible brane has a negative
tension and the bulk cosmological constant has to be negative as in the RS 
case.  
\\
\indent
Unless the input parameters are finetuned like the Eq.~(\ref{gbfine}), 
the branes
and the bulk space are not static any more\cite{infla}. Then, for inflationary 
solutions in the RS model with the Gauss-Bonnet term, we assume a separable 
metric ansatz 
like $n=f(y)$, $a=f(y)g(\tau)$ in the Eq.~(\ref{metric}). 
Here we have the extra dimension static necessarily for the separable metric; 
$b=b_0=const$ and $\dot{g}/g=H_0=const$. 
Consequently,  
the inflationary solutions are two-fold as follows\cite{KKL},
\begin{eqnarray}
ds^2=\bigg(\frac{H_0}{k_\pm}\bigg)^2 sinh^2(-k_\pm b_0|y|+c_0)[-d\tau^2
+e^{2H_0\tau}\delta_{ij}dx^i dx^j]+b^2_0 dy^2
\end{eqnarray}
where the constants $b_0$ and $c_0$ are determined from the boundary conditions
on the branes. In the limit of $H_0\rightarrow 0$ and $c_0\rightarrow +\infty$ 
with keeping the ratio $(H_0 e^{c_0})/(2k_\pm)\rightarrow 1$ fixed, the two RS 
type static solutions are recovered along with the consistency from the 
boundary conditions, Eq.~(\ref{gbfine}). Therefore, one can see the possibility 
of the visible brane with the positive tension again.  
By making the 4-dimensional part of the metric be in the form 
$ds_{4}^2=-dt^2+e^{2H(y)t}\delta_{ij}dx^idx^j$, we get the Hubble parameter 
at the visible brane expressed as 
$H_{\rm vis,\pm}=\sqrt{ (k_{\rm vis,\pm})^2 -k_{\pm}^2}$. 
Here $k_{\rm vis,\pm}^2=k^2_\pm$ for the static
solutions and the two parameters corresponding to the $k_+$ and $k_-$ solutions 
at the visible brane, $k_{\rm vis,\pm}$, are given by
\begin{equation} 
k_{\rm vis,\pm}={\left(\Lambda_2^\pm+\rho_{\rm vis}\right)\over
6M^3\sqrt{1+({4\alpha\Lambda_b}/{3M^5})}}
\end{equation}
where $\Lambda_2^\pm \gg\rho_{\rm vis}$.
Thus the Hubble parameter at the visible brane is given by 
\begin{equation}
H_{\rm vis,\pm}^2=\frac{\rho_{\rm vis}(\rho_{\rm vis}+2\Lambda_2^\pm)
}{36M^6(1+4\alpha\Lambda_b/3M^5)}
=\frac{\pm \rho_{\rm vis}}{3M_{P}^2\sqrt{1+4\alpha\Lambda_b/3M^5}}
\left[1+\frac{\rho_{\rm vis}}{2\Lambda_2^\pm}\right]~~.
\end{equation}
Therefore, with the $k_+$ solution we can obtain a plausible FRW universe at low temperatures. As a result, our additional solution 
could be proposed to solve the negative tension problem in the RS model. 
However, as we will see in subsequent sections, we will show that the $k_+$ 
solution may be unstable under perturbations.

\section{RS model with the Gauss-Bonnet term as a 4D gravity theory}
In the second RS model with a single brane of positive tension\cite{RSII}, it 
has been 
shown that gravity can be localized on the brane even if the extra dimension 
is non-compact. As a result, the 4D Newtonian gravity can be reproduced on the 
brane without the need of compactifying the extra dimension.  
The 4D graviton is identified as a normalizable bound state of 
massless graviton due to the delta function source of the brane and 
continuous Kaluza-Klein modes give rise to small corrections to the 4D 
Newtonian gravity since they are weakly coupled to the brane matters\cite{RSII}.  However, it seems that it is 
not plausible to detect the extra dimension in the second RS model because
the effects from the extra dimension appear around the AdS curvature scale $k$, 
which may be about the Planck scale for giving no hierarchy.
(There also exists a stringy picture of lowering the AdS curvature 
scale.\cite{rsexp}) 
On the other hand, the localization of gravity has
been also shown by decomposing the full graviton propagator\cite{GT,GKR}. 
As a result, it 
turns out that a localized source induces a localized field, which diminishes 
as one goes toward the AdS horizon\cite{GT,GKR}. And, the brane bending effect 
in the 
existence of matter on the brane is shown to be crucial for consistency of 
the linearized approximation\cite{GKR} and is necessary to reproduce 
the 4D Einstein gravity on the brane\cite{GT,GKR}. 
\\
\indent
For the second RS model, the extra dimension is non-compact with 
$y\in(-\infty,\infty)$, of which just the half $[0,\infty)$ is sufficient for
discussion. Having the perturbed metric as $g_{MN}=\bar{g}_{MN}+h_{MN}$, 
Randall and 
Sundrum\cite{RSII} took the gauge of $h_{55}=h_{5\mu}=0$ (Gaussian normal 
condition) and 
$\partial^\mu h_{\mu\nu}=
h^\mu_\mu=0$ (transverse traceless condition) in the absence of matter on the brane, of which the advantage 
is that all
components of the metric are decoupled. In general, however, the metric does 
not satisfy the RS gauge condition on the brane with matter and thus we
have to maintain some degrees of freedom of the metric to satisfy the brane
junction condition. As a result, there exists an additional unphysical scalar
degree of freedom, which is harmful because it might couple to the trace of 
the energy-momentum tensor. However, it has been shown that the scalar degree 
can be cancelled out by a fifth coordinate transformation (or brane bending)
\cite{GT}. 
\\
\indent
For the case in the second RS model with the Gauss-Bonnet term\cite{KKL,KL},  
we choose just the 
Gaussian normal condition for the metric perturbation for the case with matter
on the brane. Here we put $b_0=1$ in the Eq.~(\ref{rsm}) and assume that the
matter is localized on the brane, i.e., $T_{55}=T_{5\mu}=0$ and 
$T_{\mu\nu}=S_{\mu\nu}(x)\delta(y)$. (Note that $T_{\mu\nu}$ are not including 
the contribution from the brane tension.)    
Then, the equation of motion for the trace $h$ follows\cite{KL}, 
\begin{eqnarray}
&\partial_y\bigg[e^{-2ky}\partial_y (e^{2ky}h)\bigg]=\frac{2}{3}M^{-3}
\bigg(1-\frac{4\alpha k^2}{M^2}\bigg)^{-1}T^\mu\,_\mu.\label{eqh}
\end{eqnarray} 
Therefore, if $T^\mu\,_\mu\neq 0$, the trace $h$ has the exponentially growing
component. So, to cancel the growing component for validity of the linearized
approximation, we have to take the $y$ position of the brane shifted 
by $-\xi^5$, 
\begin{eqnarray}
&\partial^\mu\partial_\mu\xi^5(x)=\frac{1}{6}M^{-3}\bigg(1-\frac{4\alpha k^2}
{M^2}\bigg)^{-1} S^\mu\,_\mu\label{bend}
\end{eqnarray}
In fact, $\xi^5$ is the gauge choice of the 5D coordinate transformation 
maintaining the metric as a Gaussian normal form.
Then, we can always choose the transverse traceless condition (i.e., the RS 
gauge) for the metric
by rewriting the Eq.~(\ref{eqh}) and the relation 
$\partial_y (e^{2ky}\partial^\lambda h_{\mu\lambda})=
\partial(e^{2ky}\partial_\mu h)$ in the coordiate where the brane is 
shifted by $-\xi^5$ along the bulk. As a result, the brane bending $\xi^5$ 
will play the role of the source for the metric perturbation in the RS gauge. 
Consequently, in the initial coordinate where the brane is perpendicular to the
AdS horizon, the metric perturbation on the brane is made of two components 
as follows\cite{KL},
\begin{eqnarray}
h_{\mu\nu}(x)=h^{(m)}_{\mu\nu}(x)+h^{(b)}_{\mu\nu}(x)\label{pertunprime}
\end{eqnarray}
where
\begin{eqnarray}
h^{(m)}_{\mu\nu}(x)&=&-M^{-3}\int d^4 x^\prime G_5(x,0;x^\prime,0)
\bigg(S_{\mu\nu}(x^\prime)-\frac{1}{3}
\eta_{\mu\nu}S^\lambda_\lambda(x^\prime)\bigg)\label{matter}
\\
h^{(b)}_{\mu\nu}(x)&=&2k\eta_{\mu\nu}\xi^5(x)\label{bending}
\end{eqnarray}
where $G_5$ is the 5D graviton propagator in the presence of the Gauss-Bonnet
term\cite{KL}. For instance, in case of a 
static point source with mass $m$ on the brane, i.e., for the 
energy-momentum tensor $S_{\mu\nu}=m\delta_{\mu 0}\delta_{\nu 0}
\delta^{(3)}(x)$, we obtain the approximate metric perturbation for a static
point source on the brane as the following\cite{KL},
\begin{eqnarray}
h_{00}(x)&=&\frac{2G_N m}{r}\bigg[1+\frac{2}{3}
\bigg(1-\frac{2}{3}\beta\bigg)^{-1}\bigg(\frac{1}{1+2\beta}\bigg)^2
\frac{1}{(kr)^2}\bigg],\label{fm00} \\
h_{ij}(x)&=&\frac{2G_N m}{r}\bigg[\bigg(\frac{1+\frac{2}{3}\beta}
{1-\frac{2}{3}\beta}\bigg)+\frac{1}{3}\bigg(1-\frac{2}{3}
\beta\bigg)^{-1}\bigg(\frac{1}{1+2\beta}\bigg)^2\frac{1}{(kr)^2}
\bigg]\delta_{ij}, \label{fmij}
\end{eqnarray}
where by the Newton potential $\Phi_N=-\frac{1}{2}h_{00}$, the Newton 
constant is given by 
\begin{eqnarray}
G_N &\equiv& \frac{k}{8\pi M^3}\bigg(1-\frac{4\alpha k^2}{M^2}\bigg)^{-1}
\bigg(\frac{1-\frac{2}{3}\beta}{1+2\beta}\bigg), \\
\beta&\equiv& \frac{4\alpha k^2/M^2}{1-12\alpha k^2/M^2}.
\end{eqnarray}
For the $k_+$ solution, the Newton constant $G_N$ would be negative and 
$1-\frac{2}{3}\beta>0$ always, which might give rise to massless and massive
ghosts. That is, it means that the $k_+$ solution is unstable under 
perturbations and therefore we have to exclude it at the perturbative level.
On the other hand, for the $k_-$ solution, we can get the normal 
gravity without ghosts for $-0.47<\frac{4\alpha \Lambda_b}{3M^5}\leq 0$ 
for $\alpha>0$ or always for $\alpha<0$. Therefore, even the $k_-$ 
solution could excite ghost particles in some bulk parameter space 
with $\alpha>0$. 
\\
\indent
As leading terms of $h_{00}$ and $h_{ij}$ components are not equal in the above 
result, elimination of the unphysical scalar degree due to the 
brane bending effect is incomplete in the presence of the Gauss-Bonnet term,
unlike in the original RS case.
Therefore, we can also show that the bending of light passing by the 
Sun could be modified with the Gauss-Bonnet term. For a source 
in $xz$ plane on the brane, for instance, the bending of light 
travelling in $z$ direction is described by a Newton-like force law, 
$\ddot{x}=\frac{1}{2}(h_{00}+h_{zz})_{,x}$. If the metric perturbations 
due to the Sun are approximated by those from a point source, 
the bending of light is $(1-\frac{2}{3}\beta)^{-1}$ of 
that predicted from the 4D Einstein gravity. Therefore,  
from the experimental measurements of the bending of light\cite{will}, 
we can get another bound on 
the Gauss-Bonnet coefficient as $-0.20<\frac{4\alpha \Lambda_b}{3M^5}
<1.2$ for the $k_-$ solution connected to the RS solution.

\section{Conclusions}
We studied static and inflationary solutions in the Randall-Sundrum framework
with the Gauss-Bonnet term added to the standard Einstein term. It has been 
argued that the Gauss-Bonnet term is the only consistent curvature squared 
interaction in the Randall-Sundrum model. In particular,  
our additional RS type solution might solve the negative tension problem 
but it may be unstable under perturbations. And  we showed that for a 
vanishing Gauss-Bonnet coefficient, the brane
bending allows us to reproduce the 4D Einstein gravity at the linearized level.

\section*{Acknowledgments}
The author thanks S.-H. Moon and J. D. Park for useful discussions. 
This work is supported by the BK21 program of Ministry of Education, Korea.

\end{document}